\documentclass{iaus}
\usepackage{graphicx}

\title[IAUS265.~~Stellar photospheric parameters and C abundances]
{Photospheric parameters and C abundances \\
in solar-like stars with and without planets}

\author[Ronaldo Da Silva \& Andr\'e Milone]
{Ronaldo Da Silva$^1$ \and Andr\'e Milone$^1$}

\affiliation{$^1$Divis\~ao de Astrof\'\i sica, Instituto Nacional de
Pesquisas Espaciais, Brazil}

\pubyear{2009}
\volume{265}
\pagerange{1--2}
\jname{Chemical Abundances in the Universe: Connecting First Stars to Planets}
\editors{K. Cunha, M. Spite \& B. Barbuy, eds.}
\begin{document}

\maketitle

%
%
\begin{abstract}
We have been analyzing a large sample of solar-like stars with and without
planets in order to homogeneously measure their photospheric parameters and
Carbon abundances. Our sample contains around 200 stars in the solar
neighborhood observed with the ELODIE spectrograph, for which the
observational data are publicly available. We performed spectral synthesis
of prominent bands of C$_{2}$ and C~I lines, aiming to accurately obtain the
C abundances. We intend to contribute homogeneous results to studies that
compare elemental abundances in stars with and without known planets. New
arguments will be brought forward to the discussion of possible chemical
anomalies that have been suggested in the literature, leading us to a better
understanding of the planetary formation process. In this work we focus on
the C abundances in both stellar groups of our sample.
\end{abstract}

%
%
\section{Introduction}

The Sun is widely thought to be formed from material representative of local
physical conditions in the Galaxy at the time of its formation and to
represent a standard chemical composition. This {\it homogeneity hypothesis}
has been often put in question because of many improvements in the
observations techniques and data analysis. With the discovery of new
planetary systems, the already known {\it heterogeneity} sources (stellar
nucleosynthesis, stellar formation process) have gained a new perspective
and brought new questions. It is now a fact that stars with giant planets
are, on average, richer in metals than those for which no planet was
detected (\cite[Gonzalez 2006]{Gonzalez2006}
and references therein). Some authors
suggested that this kind of anomaly may not only involve the content of
heavy elements but also some light elements like Li, C, N, and O
(\cite[Ecuvillon et al. 2006]{Ecuvillonetal2006}
and references therein). Other authors found no
difference in the abundance of light elements when comparing stars with and
without planets (see \cite[Ecuvillon et al. 2004]{Ecuvillonetal2004}
and references therein). The
studies above are not conclusive yet. We need new tests, using more accurate
and homogeneous data, with a larger number of stars. Abundances of these
elements in solar-like stars will bring new information that will surely
help to distinguish the different stellar and planetary formation processes.
This is the purpose of our work, in which we analyzed a sample of 200 stars
to homogeneously measure the photospheric parameters and C abundances.

%
%
\section{Data and Method}

Our sample consists of 200 single solar-like stars in the solar
neighborhood, observed in high signal-to-noise ratio ($\ge$~200) and high
resolution (42\,000) with the ELODIE spectrograph
(\cite[Baranne et al. 1996]{Baranneetal1996})
of the Haute Provence Observatory (France), and for which data are publicly
available (\cite[Moultaka et al. 2004]{Moultakaetal2004}).
We obtained $T_{\rm eff}$, [Fe/H],
log(g), and $\xi$ (through the excitation equilibrium of neutron iron and
the ionization equilibrium between Fe~I and Fe~II). We also performed
spectral synthesis of some regions containing bands of the ${\rm C}_2$ Swan
System (e.g. at $\lambda$5165), using the MOOG LTE code (Sneden 2000,
http://verdi.as.utexas.edu/moog.html).

%
%
\section{Results and discussion}

\begin{figure}[b]
\centering
\begin{minipage}[b]{0.42\textwidth}
\centering
\resizebox{\hsize}{!}{\includegraphics{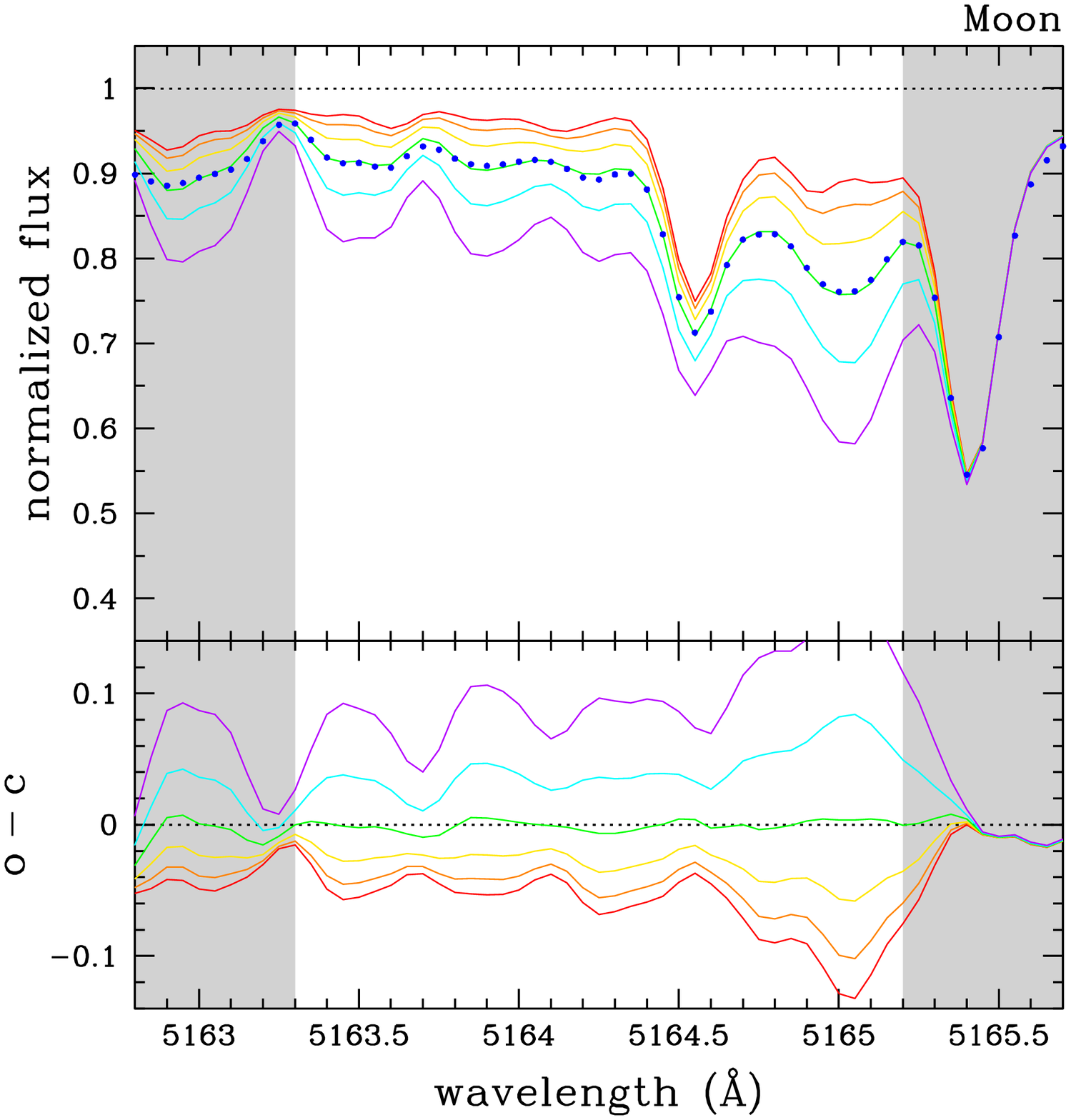}}
\end{minipage}
\begin{minipage}[b]{0.57\textwidth}
\centering
\resizebox{\hsize}{!}{\rotatebox[origin=rb]{-90}{\includegraphics{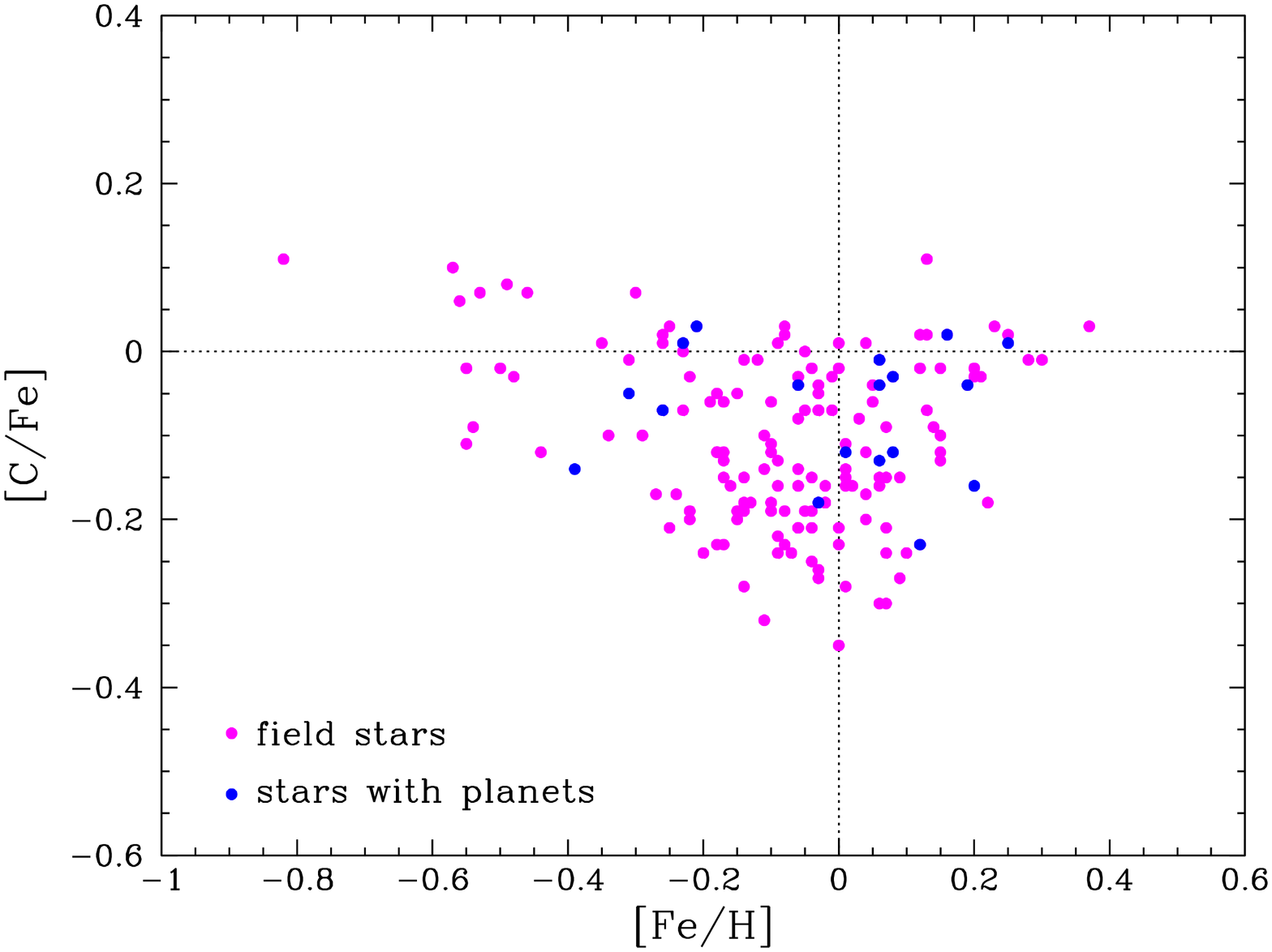}}}
\end{minipage}
\caption{{\it Left:} spectral synthesis applied to the solar spectrum for
          several C abundances (in steps of 0.1 dex). The best fit gives
	  [C/Fe] = 0.01 dex. {\it Right:} [C/Fe] {\it versus} [Fe/H] diagram
	  comparing stars with and without planets.}
\label{fig}
\end{figure}

The present work is still ongoing and we show here our first results:
photospheric parameters and C abundances based on the intensity of the
${\rm C}_2$~(0,0) band head at $\lambda$5165. A comparison of the
photospheric parameters here obtained to those published by other works
having stars in common shows a good agreement between the samples.
Figure~\ref{fig} (left) shows the solar spectrum and synthetic spectra for
different C abundances. The right panel shows the diagram [C/Fe] versus
[Fe/H], comparing stars with and without planets. It seems that stars with
planets are slightly richer in C than field stars, but a larger number of
planetary systems is required to confirm this possibility.

%
%
\section{Conclusions}

The preliminary results on C abundances are presented here for a large
number of nearby solar-like stars, based on homogeneous photospheric
parameters obtained from spectra with high signal-to-noise ratio and high
resolution. Our analysis used public spectra from the ELODIE database, which
represent about 90\% of all data. The remaining 10\% include many stars with
detected planets and having many observations that shall be analyzed in the
same way as soon as they become available to the scientific community, since
they will contribute to more reliable conclusions.


\end{document}